\documentclass[twocolumn]{aastex631}
\usepackage{showyourwork}

\newcommand{\sbu}{Department of Physics and Astronomy, Stony Brook University, Stony Brook NY 11794, USA}
\newcommand{\cca}{Center for Computational Astrophysics, Flatiron Institute, New York NY 10010, USA}

\usepackage{bbold}

\usepackage[commandnameprefix=always,commentmarkup=uwave]{changes}
\definechangesauthor[name={Sabina}, color=cyan]{S}
\definechangesauthor[name={Will}, color=magenta]{W}
\definechangesauthor[name={Brett}, color=yellow]{B}
\definechangesauthor[name={Rodrigo}, color=green]{R}

\usepackage{hyperref}

\newcommand{\paperone}{\hyperlink{cite.paper1}{Paper~I}}

\usepackage{fontawesome5}  

\begin{document}

\title{Polka-dotted Stars II: Starspots and obliquities of Kepler-17 and Kepler-63}

\author[0000-0002-6650-3829]{Sabina Sagynbayeva}
\email{sabina.sagynbayeva@stonybrook.edu}
\affiliation{\sbu}

\author[0000-0003-1540-8562]{Will M. Farr}
\affiliation{\sbu}
\affiliation{\cca}




\begin{abstract}
Starspots trace stellar magnetic activity and affect both stellar evolution and exoplanet
characterization. While occultation-based spot studies have been applied to individual systems,
comparative analyses remain limited. We apply the \texttt{StarryStarryProcess} Bayesian surface
mapping framework to archival \textit{Kepler} light curves of two planet hosts—Kepler-63 and Kepler-17—extending 
the validation established on TOI-3884 (Paper~I).
Across both systems we infer characteristic spot radii of $<$10$^\circ$.
The latitudinal distributions of both G-dwarfs exhibit active latitudes (Kepler-63 at $\pm$30$^\circ$, Kepler-17 at
$\pm$15$^\circ$). Our analysis yields stellar obliquity measurements in excellent agreement with previous
studies for both systems. This validates our methodology and demonstrates that transit-based
surface mapping can simultaneously recover planetary parameters, stellar orientations, and
magnetic morphologies. The analysis of the systems inferred nearly aligned (Kepler-17) to highly
misaligned (Kepler-63) geometries.

\end{abstract}

\section{Introduction}
\label{sec:intro}

Starspots are key tracers of stellar magnetic activity, shaping a variety of observable behaviors 
including rotational modulation, wavelength-dependent flux variations, 
and signatures during planetary transits. These surface features influence both stellar physics and exoplanet 
characterization, where they can bias 
radius measurements or imitate atmospheric signatures -- what we call a ``stellar contamination" in transmission spectroscopy. 
The need to accurately account for these 
effects has grown with the emergence of high-precision time-series data, as demonstrated by works
of active exoplanet hosts across broadband photometry and infrared 
spectroscopy \citep[e.g.,][]{Ducrot2018, Morris2018,Sanchis2011,Valio2017}.

Recent observations have further emphasized the challenges posed by stellar contamination. In several JWST datasets, 
small-scale surface heterogeneities have been shown to interfere with atmospheric retrievals at levels approaching 
the precision of the instrument \citep[e.g.,][]{Lim2023}. At the same time, the long and continuous light curves 
from missions such as \textit{Kepler} \citep{Borucki2010} and \textit{TESS} contain a wealth of information about the distribution, 
evolution, and stability of starspots. Rotational modulation offers global constraints on longitudinal structure, 
while planetary transits provide direct, localized probes of the stellar surface. 
The basic idea of inferring surface features from transit bumps dates back to the work of 
\cite{Silva2003} and has since inspired a broad set of methods for mapping spot properties from occultation 
signals \citep[e.g.,][]{Huber2010, Oshagh2013, Tregloan-Reed2013, Beky2014, Montalto2014, Maxted2016, Herrero2016, Juvan2018, Scandariato2017,paper1}.

In \cite{paper1} (hereafter \paperone), we introduced the \texttt{StarryStarryProcess} framework as a flexible, 
probabilistic model for interpreting these combined signatures of stellar activity. By coupling the spherical  
harmonic machinery of \texttt{starry} \citep{Luger2019} with the spot-population formalism 
of \texttt{StarryProcess} \citep{Luger2021b}, the method treats rotational variability and 
spot-crossing events within a unified generative model. This approach was shown to alleviate the 
classical degeneracies that arise when interpreting rotational modulation alone \citep{Luger2021a}. 
Applied to the young, active system TOI-3884, the framework demonstrated that joint modeling can 
simultaneously constrain the stellar inclination, obliquity, and statistical properties of the spot distribution.

Occultation-based spot coverage studies have primarily focused on individual systems or limited
epochs of high-quality data. With the wealth of archival \textit{Kepler} photometry, however, we
now have the opportunity to extend such analyses to long-baseline, high-precision light curves
of multiple active host stars. A natural extension of this work is to test the methodology across stars 
with different magnetic morphologies and 
levels of activity. The four-year, high-cadence photometry 
from \textit{Kepler} is exceptionally well suited for such a comparative study of different exoplanet hosts
that exhibit different magnetic activity levels. \textit{Kepler}'s continuity allows spot evolution to be followed 
over many rotations and numerous transits, enabling the study of active latitudes and stellar orientation. 
Two systems, Kepler-63 and Kepler-17, stand out as prime targets 
due to their strong rotational modulation and frequent spot occultations, making them excellent laboratories 
for exploring how magnetic activity manifests across different stellar hosts.

In this work, we apply \texttt{StarryStarryProcess} to both Kepler-63 and Kepler-17 in order to  
extract starspot population properties in a consistent manner, evaluate how our inferences compare with 
previous transit-based analyses, and derive updated constraints on the stellar obliquities. 
By analyzing these two systems using a common probabilistic framework, we extend the validation of 
\paperone to a broader astrophysical setting and assess how well the method captures real diversity in 
spotted surfaces. More generally, this work illustrates how long-baseline transit photometry can 
be used not only to characterize individual stars, but also to build a comparative picture of 
magnetic environments across exoplanet hosts.

The structure of the paper is as follows. Section~\ref{sec:model} describes the 
modeling framework and inference procedure. Section~\ref{sec:kepler63} presents 
the analysis of Kepler-63, followed by the results for Kepler-17 in Section~\ref{sec:kepler17}. 
We discuss limitations and implications in Section~\ref{sec:discussion}.

\section{Methodology}
\label{sec:model}
We employed the \texttt{StarryStarryProcess} \citep[][]{paper1} Bayesian model to map the surface features of 
two stars hosting transiting planets: Kepler-63 and Kepler-17. The \texttt{StarryStarryProcess} 
framework utilizes Bayesian inference to reconstruct stellar surface maps by analyzing transit light curves, 
leveraging the rotational modulation and transit signatures to constrain surface brightness variations. 
The model incorporates prior knowledge of stellar properties and transit parameters while accounting for 
observational uncertainties through a probabilistic approach.

To account for the temporal evolution of stellar surface features, we divided the photometric time series for 
each target into discrete chunks of duration $P_{\rm rot}/2$, where $P_{\rm rot}$ is the stellar rotation period. 
This chunking approach assumes that starspots evolve on timescales comparable to or shorter than the stellar 
rotation period, and allows us to capture temporal variations in the surface magnetic field topology while 
maintaining sufficient photometric coverage within each chunk for robust surface reconstruction. 
The analysis was performed on an ensemble of these chunks, with the reported surface maps representing 
the averaged configuration over each chunk period. This methodology enables us to track the evolution of 
magnetic features while avoiding the assumption of static surface patterns over extended observing periods.

In \paperone, we introduced and tested two ways of handling the temporal evolution of spots: 
one in which consecutive surface maps were linearly interpolated to enforce smooth temporal evolution, 
and another in which each chunk was treated independently. Both approaches produced consistent qualitative behavior, 
but the interpolation-based method introduced stronger correlations between adjacent chunks, 
making the sampling procedure substantially more challenging. In principle, the most rigorous way to model smooth 
temporal evolution would be to impose a Gaussian Process (GP) kernel in time on the spherical-harmonic coefficients 
themselves \citep{Luger2021b}. However, coupling a GP temporal prior with a high–degree spherical-harmonic expansion leads to a 
large parameter space and an inference problem that is not yet computationally tractable for our datasets. 
For these reasons, in the present work we adopt the simpler non-interpolated, independent-chunk approach, 
while planning to explore GP-regularized temporal evolution in future work.

The model accounts for the obliquity-inclination degeneracy discussed in \paperone, where the stellar obliquity ($\psi$) 
and stellar inclination ($i_{\star}$) can be degenerate between northern and southern hemispheres. By incorporating 
both rotational modulation and transit constraints simultaneously, our approach breaks these degeneracies 
to provide robust estimates of the three-dimensional stellar orientation and surface feature distributions.

Rather than reporting the traditional model parameters of contrast ($\mathbb{c}$)
and number of spots ($\mathbb{n}$) as in \paperone, we focus on physically measurable quantities 
that directly relate to observable stellar variability. The $\mathbb{n}$ parameter is not truly a spot count but 
rather a scaling parameter on the mean and variance of the spot flux decrement, and therefore does not naturally 
correspond to the number of spots visible in surface maps. Similarly, the contrast parameter $\mathbb{c}$ does not directly 
track the average contrast of individual spots. Instead, we characterize stellar activity through two measurable 
parameters: the average fractional decrement in stellar flux due to spots, 
$\mathbb{d} = \mathbb{n} \cdot \mathbb{c} \cdot \mathbb{r}^2/R_{\star}^2$, 
which depends on the combination of all three model parameters and represents the observable flux reduction; and the 
root-mean-square (RMS) fractional fluctuation in stellar flux decrement due to spots, 
$\rm{RMS}_{\rm spot} = \mathbb{d} / \sqrt{n}$, which quantifies the amplitude of rotational modulation 
caused by the inhomogeneous spot distribution.

For each target star, we processed the available photometric time series data, extracting transit events and 
out-of-transit rotational modulation to inform the surface reconstruction within each temporal chunk. We also binned the out-of-transit portion of the light curves.
The Bayesian framework allows for proper propagation of uncertainties throughout the analysis, 
providing posterior distributions for all derived parameters including spot properties, stellar orientation, 
and surface brightness maps.

\section{Kepler-63}
\label{sec:kepler63}
Kepler-63 is a G-type main sequence star hosting Kepler-63b, a sub-Neptune that orbits every 9.434 days at a 
distance of 0.08 AU. With a mass of 0.38 $M_J$ (approximately 120 Earth masses) and a radius of 0.55 $R_J$ (6.1 Earth radii), 
the planet has a density of 2.01 g/cm$^3$. This system is particularly notable for its high stellar obliquity and active 
stellar surface.

\begin{figure*}[hbt!]
    \begin{centering}
        \includegraphics[width=\linewidth]{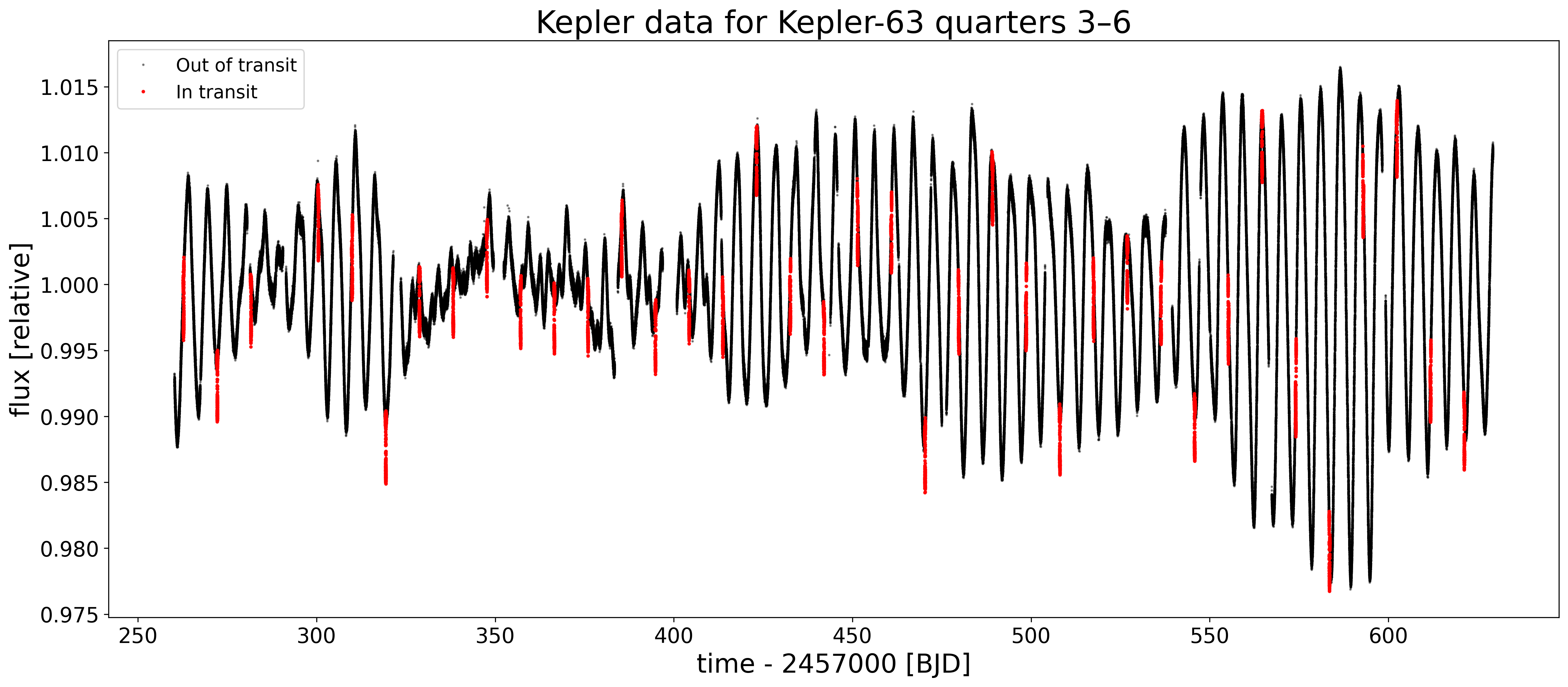}
        \caption{Light curve of Kepler-63 showing relative flux measurements over time for quarters 3, 4, 5, and 6 of 
        the Kepler mission. The red points show the transits.
        The data spans approximately 350 days (from BJD 2457250 to 2457600) and shows periodic 
        transit events appearing as characteristic dips in the stellar brightness against the background of rotational modulation from starspots.}
        \label{fig:kepler63-pdscap-lc}
    \end{centering}
\end{figure*}

The Kepler light curve for this system spans multiple quarters and reveals both the periodic transit signature and 
the quasi-periodic rotational modulation caused by evolving starspots. Figure \ref{fig:kepler63-pdscap-lc} presents the 
complete photometric time series from quarters 3-6, covering approximately 350 days. The characteristic transit dips 
are clearly visible against the background of stellar variability, which reaches amplitudes of several percent due 
to the high level of magnetic activity on this young star.

\begin{figure*}[hbt!]
    \begin{centering}
        \includegraphics[width=\linewidth]{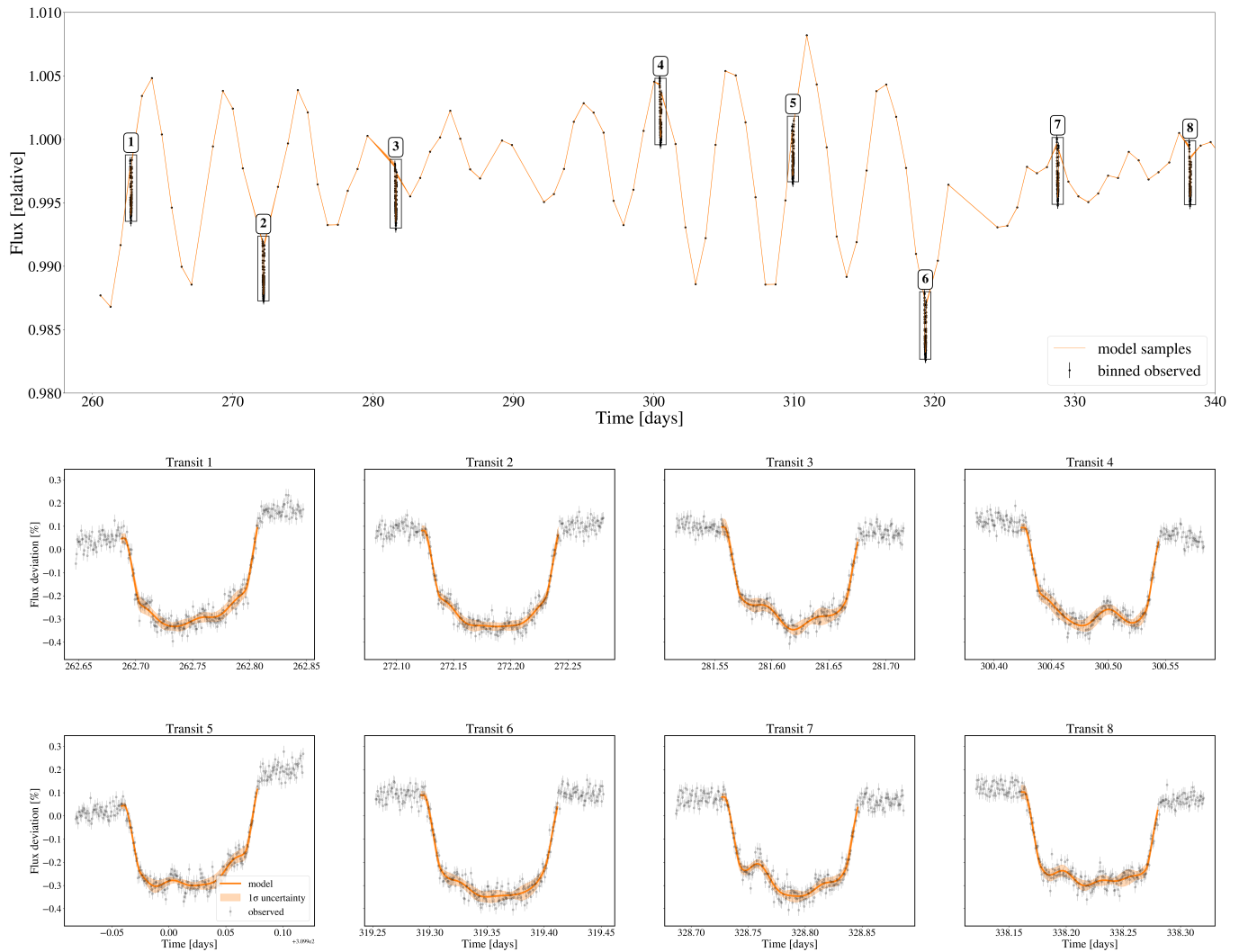}
        \caption{Transit photometry and model fits for Kepler-63. Top: Complete light curve showing eight 
        detected transits over 80 days of observations. Black points show binned detrended photometry; orange lines represent 
        model averages of 200 samples. The numbers above the boxes indicate the transit number to match the bottom panel. 
        Bottom panels: Individual transit events with observed 
        photometry (grey points with error bars) and the average model across 200 samples (orange lines) with $1\sigma$ uncertainty. 
        Model parameters were determined through simultaneous fitting of all transits and the rotational modulation.}
        \label{fig:kepler63-results}
    \end{centering}
\end{figure*}

To characterize both the planetary parameters and stellar activity, we applied our joint modeling approach 
to a representative 80-day segment of the light curve. Figure \ref{fig:kepler63-results} shows the results of 
this analysis, with the upper panel displaying the complete time series and the lower panels showing individual transit fits. 
The orange curve in the upper panel represents the detrended photometry, while black points show the binned model predictions. 
Each individual transit (grey points with error bars) is well-fitted by our model (purple lines), with the purple uncertainty 
bands representing the $1\sigma$ model uncertainty across 200 MCMC samples. The consistency of transit depths and shapes 
across all epochs confirms the robustness of our planetary parameter determination despite the stellar activity.

\begin{figure*}[hbt!]
    \begin{centering}
        \includegraphics[width=\linewidth]{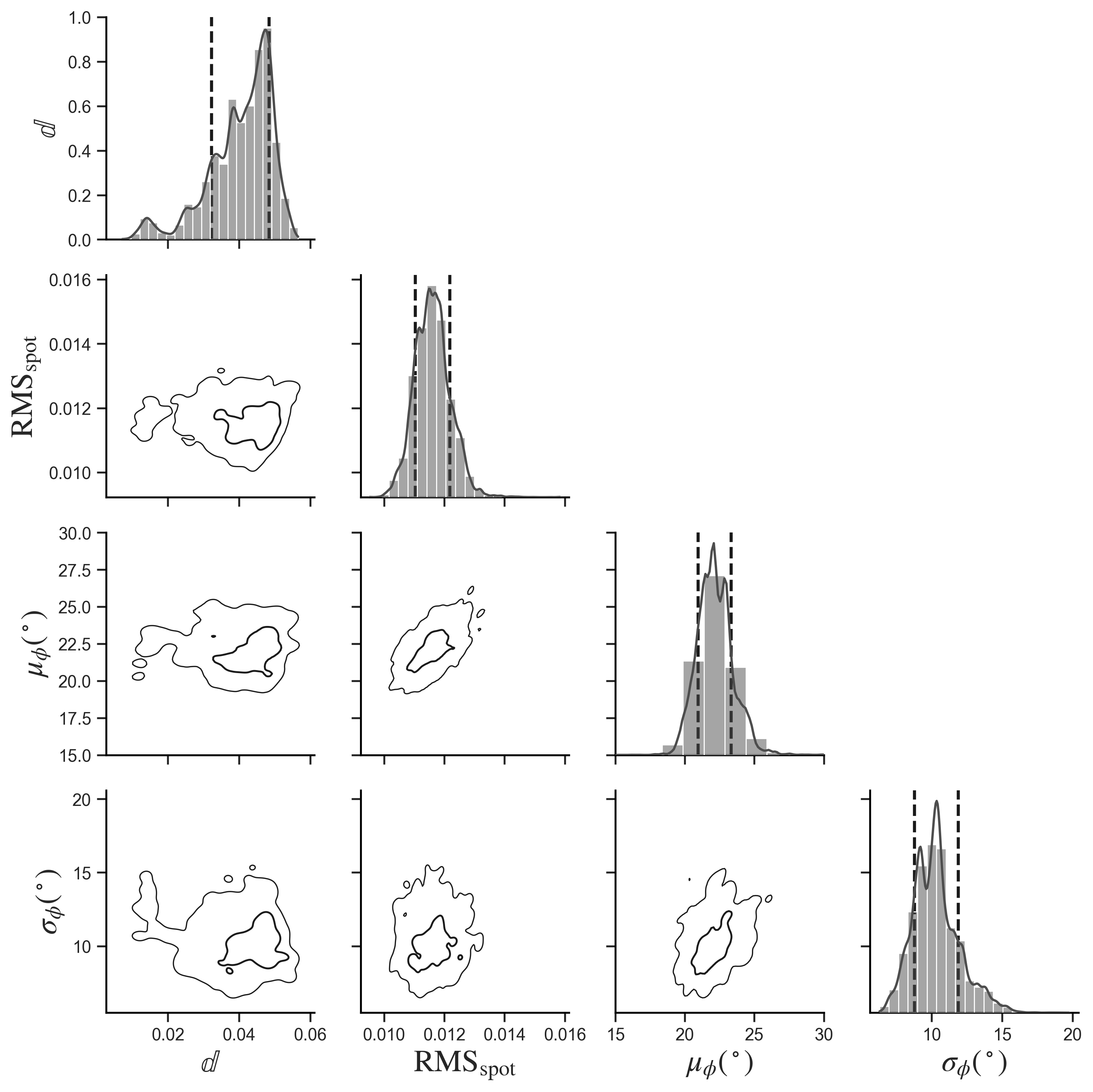}
        \caption{Corner plot showing posterior distributions and parameter correlations from our analysis of stellar 
        spot properties for Kepler-63. Diagonal panels display marginal posterior distributions for the 
        flux decrement due to spots ($\mathbb{d}$), $\rm{RMS}_{\rm{spot}}$, and mean and standard deviation of 
        spot latitudes ($\mu_\phi$ and $\sigma_\phi$). Diagonal panels show marginalized posterior distributions; 
        dashed vertical lines mark the 16th and 84th percentiles ($1\sigma$ credible intervals). 
        Off-diagonal panels show two-dimensional kernel density estimates with three iso-density contours 
        illustrating correlations between parameters. 
        The tight correlations between parameters demonstrate 
        well-constrained solutions from the photometric modeling.}
        \label{fig:kepler63-corener-gps}
    \end{centering}
\end{figure*}

Our modeling approach simultaneously constrains both the planetary system parameters and the 
properties of the stellar magnetic activity. Figure \ref{fig:kepler63-corener-gps} presents the posterior 
distributions for the key starspot parameters derived from our analysis. The corner plot reveals well-constrained 
solutions for spot radius ($\mathbb{r}$), flux decrement ($\mathbb{d}$), $\rm{RMS}_{\rm{spot}}$, and 
the mean and standard deviation of spot latitudes ($\mu_\phi$ and $\sigma_\phi$). The tight parameter correlations 
visible in the off-diagonal panels demonstrate that the photometric data contains sufficient information to 
simultaneously characterize multiple aspects of the stellar magnetic field geometry.

Our derived spot characteristics are consistent with previous studies of this system. \cite{Sanchis2013} 
reported typical spot radii of $\sim10^\circ$ with maximum sizes reaching $15^\circ-20^\circ$, which aligns 
well with our inferred spot radius of $10.02^{+0.03}_{-0.02}$ degrees (see Table \ref{tab:ResultsKepler63}). 
However, we note that our inferred radii are larger than they ought to be, and the posterior for the size of spots
hits the limit of the prior due to the lmimited order of the largest spherical harmonic degree.
The RMS fractional fluctuation in stellar flux decrement, 
$\rm{RMS}_{\rm spot}$, exhibits a narrower distribution centered near 0.009, reflecting the amplitude of 
rotational modulation caused by the inhomogeneous distribution of starspots across the stellar surface. The $\mathbb{d}$ 
parameter shows a distribution that peaks around 0.03 and extends up to about 0.04, indicating that starspots 
collectively reduce the stellar flux by approximately 3-4\% on average. 

\begin{figure*}[hbt!]
    \begin{centering}
        \includegraphics[width=\linewidth]{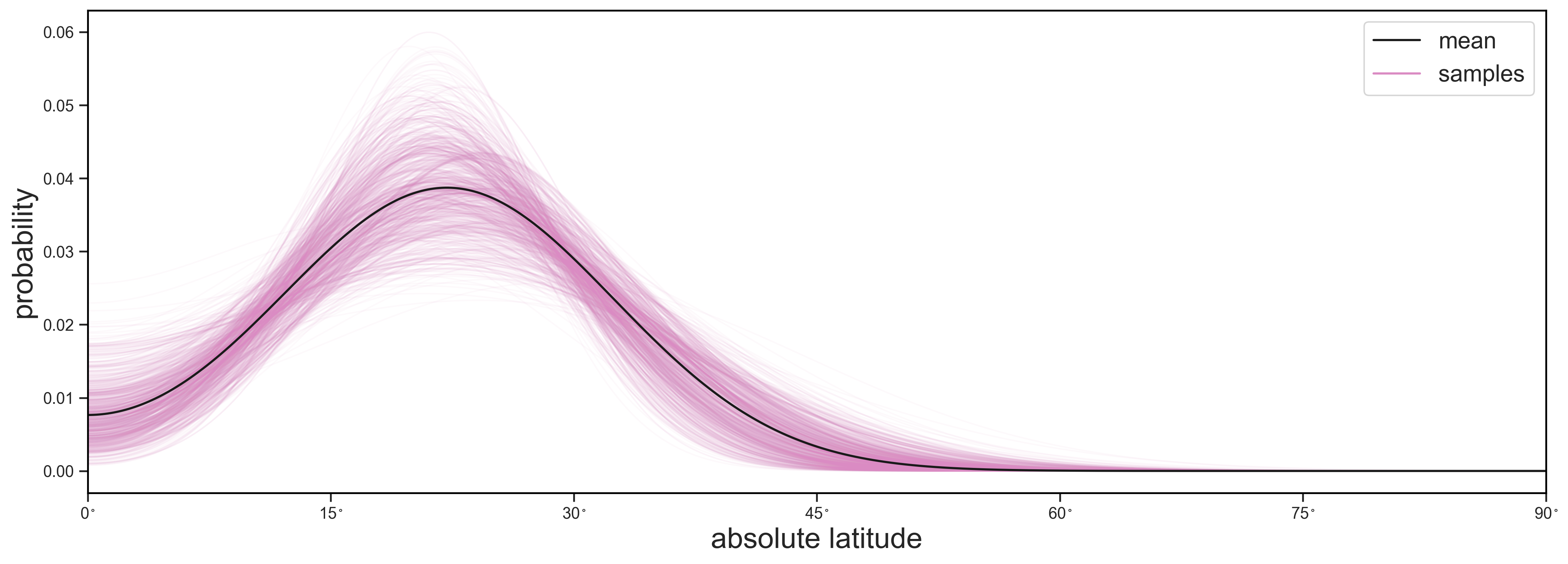}
        \caption{Latitudinal distribution of stellar activity for Kepler-63 derived from spot modeling. 
        The plot shows Posterior distributions of the absolute active latitude, shown from $0^\circ$ to $90^\circ$. 
        Thin colored curves indicate individual posterior samples, while the thick black curve shows the mean distribution.}
        \label{fig:kepler63-activelats}
    \end{centering}
\end{figure*}

Figure \ref{fig:kepler63-activelats} reveals a pronounced non-equatorial structure in the probability density 
of spot occurrence, with two distinct active latitude bands centered at approximately $\pm 30^\circ$. This pattern 
is reminiscent of solar activity cycles but occurs at higher latitudes than typically observed on the Sun. 
The black curve shows the mean posterior distribution, while the collection of pink curves represents individual 
MCMC samples, demonstrating the robustness and repeatability of this latitudinal preference. 
This level of spatial resolution in stellar activity 
mapping represents a significant advance in our ability to characterize the magnetic field structure of active stars 
through transit photometry.

\begin{figure*}[hbt!]
    \begin{centering}
        \includegraphics[width=\linewidth]{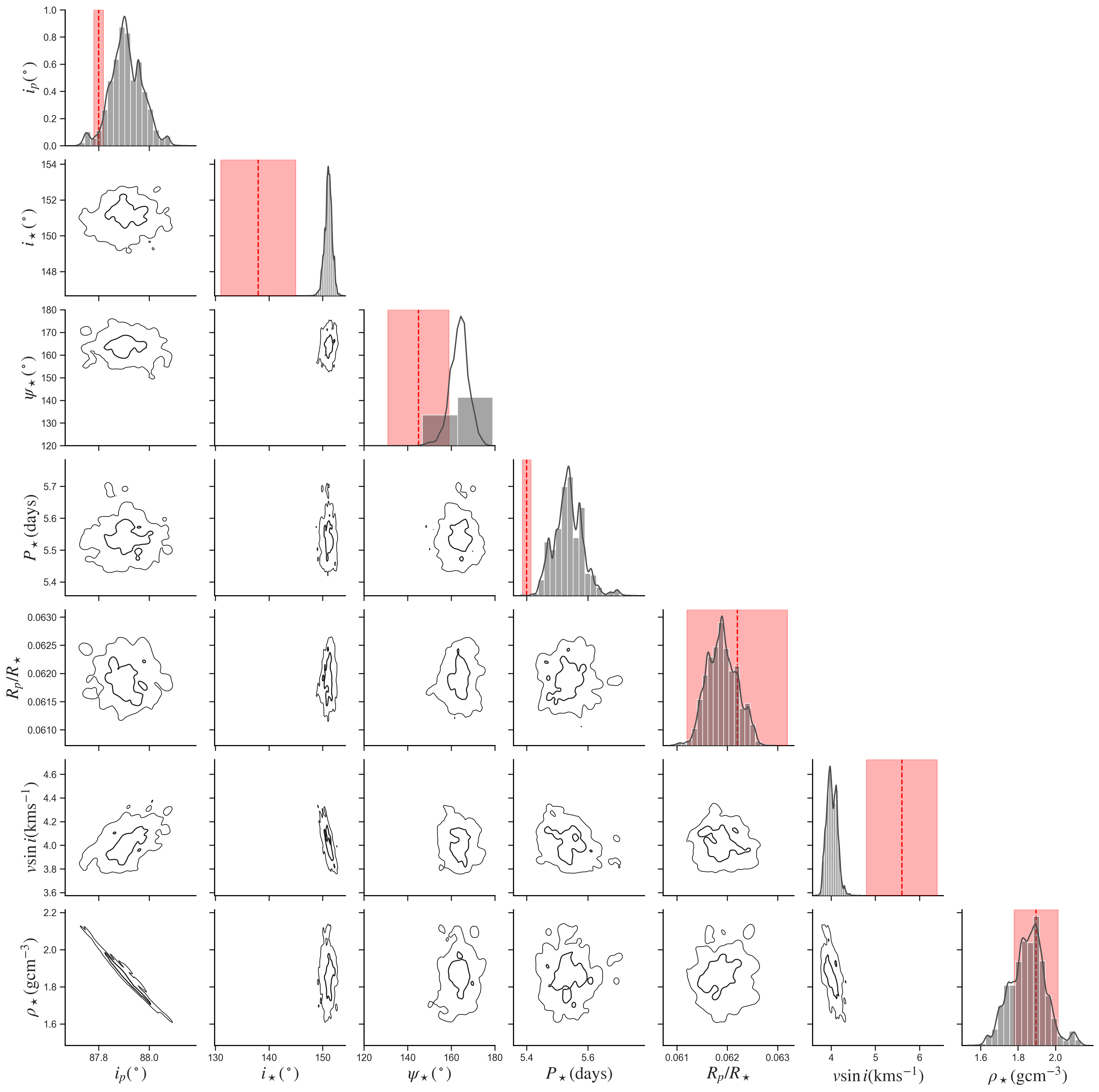}
        \caption{Corner plot showing posterior distributions and parameter correlations from our comprehensive analysis of Kepler-63. 
        Diagonal panels display marginal posterior distributions for orbital inclination ($i_p$), stellar inclination 
        ($i_\star$), stellar obliquity ($\psi_\star$), rotation period ($P_\star$), planet-to-star radius ratio ($R_p/R_\star$), 
        projected stellar rotation velocity ($v \sin{i}$), and stellar density ($\rho_\star$). 
        Off-diagonal panels show 2D posterior distributions with contour levels indicating parameter covariances. 
        Red shaded regions and dashed lines indicate literature values from \cite{Sanchis2013} for comparison. 
        The excellent agreement demonstrates the reliability of our joint modeling approach while providing enhanced precision for several key parameters.}
        \label{fig:kepler63-compare}
    \end{centering}
\end{figure*}

To validate our results, we compare our derived system parameters with the comprehensive study by 
\cite{Sanchis2013}. Figure \ref{fig:kepler63-compare} presents the joint posterior distributions for all key 
system parameters, with red shaded regions indicating the literature values for comparison. 
Table \ref{tab:ResultsKepler63} provides a detailed numerical comparison of our results with the published values. 
The agreement is generally excellent, particularly for the fundamental system properties. Our 
orbital inclination measurement of $87.9^{+0.07}_{-0.06}$ degrees is consistent with their value 
of $87.806^{+0.018}_{-0.019}$ degrees, while our planet-to-star radius ratio of $0.062 \pm 3\times10^{-4}$ agrees 
well with their determination of $0.0622\pm 0.001$.

Perhaps most importantly, our stellar obliquity constraint of $163.9^{+3.8}_{-4.2}$ degrees is consistent 
with the \cite{Sanchis2013} measurement of $145^{+9}_{-14}$ degrees, confirming the highly misaligned nature of this system. 
Our stellar rotation period of $5.53\pm 0.04$ days also agrees well with their photometric determination of 
$5.401\pm 0.014$ days. The consistency between these independent analyses strengthens confidence in both the methodology 
and the derived parameters, while demonstrating that our approach can successfully reproduce established results.

\begin{table}[]
    \vspace{0.5cm}
    \centering
    \caption{Comparison of inferred parameters for Kepler-63 between our analysis and \cite{Sanchis2013}. The table shows excellent agreement for fundamental system properties while providing new constraints on stellar magnetic activity parameters.}
    \begin{tabular}{lll}
    \hline
    Parameter                                 & This Work    & \cite{Sanchis2013} \\ \hline\hline
    $i_p (^\circ)$                            & $87.9^{+0.07}_{-0.06}$                             & $87.806^{+0.018}_{-0.019}$                    \\
    $e$                                       & $0.077^{+0.055}_{-0.053}$                          & $<0.45$                     \\
    $P (\rm days)$                            & $9.4341^{+4.4\times10^{-6}}_{-4.6\times10^{-6}}$   & $9.4341\pm 0.0000010$                   \\
    $t_0 (\rm days)$                          & $177.84\pm 2\times 10^{-4}$                        &    -                 \\
    $R_p / R_\star$                           & $0.062 \pm 3\times10^{-4}$                         &  $0.0622\pm 0.001$                   \\ 
    $u_1$                                     & $0.264 \pm 0.034$                                  &  $0.31 \pm 0.04$                   \\ 
    $u_2$                                     & $0.377^{+0.018}_{-0.044}$                          &  $0.354^{+0.07}_{-0.05}$                   \\ 
    $v\sin i (\rm kms^{-1})$                  & $4.006^{+0.12}_{-0.11}$                            & $5.6\pm 0.8$                    \\ \hline
    $i_\star (^\circ)$                        & $151.1^{+0.7}_{-0.6}$                              &  $138\pm 7$                    \\
    $\psi_\star (^\circ)$                     & $163.9^{+3.8}_{-4.2}$                              &  $145^{+9}_{-14}$                   \\
    $P_\star (\rm days)$                      & $5.53\pm 0.04$                                     & $5.401\pm 0.014$                    \\ 
    $\rho_\star (\rm gcm^{-3})$               & $1.86^{+0.08}_{-0.1}$                              & $1.89^{+0.125}_{-0.117}$                    \\ \hline
    $\mathbb{r} (^\circ)$                     & $10.02^{+0.03}_{-0.02}$                            &  -                   \\
    $\mathbb{d}$                              & $0.03^{+0.004}_{-0.007}$                           &  -                   \\
    $\rm RMS_{\rm spot}$                      & $0.008\pm0.0004$                           &  -                   \\
    $\mu_\phi (^\circ)$                       & $22.08^{+1.25}_{-1.13}$                            &  -                   \\
    $\sigma_\phi (^\circ)$                    & $10.2^{+1.69}_{-1.42}$                             &  -                   \\ \hline
    \label{tab:ResultsKepler63}
    \end{tabular}
\end{table}

\begin{figure*}[hbt!]
    \begin{centering}
        \includegraphics[width=\linewidth]{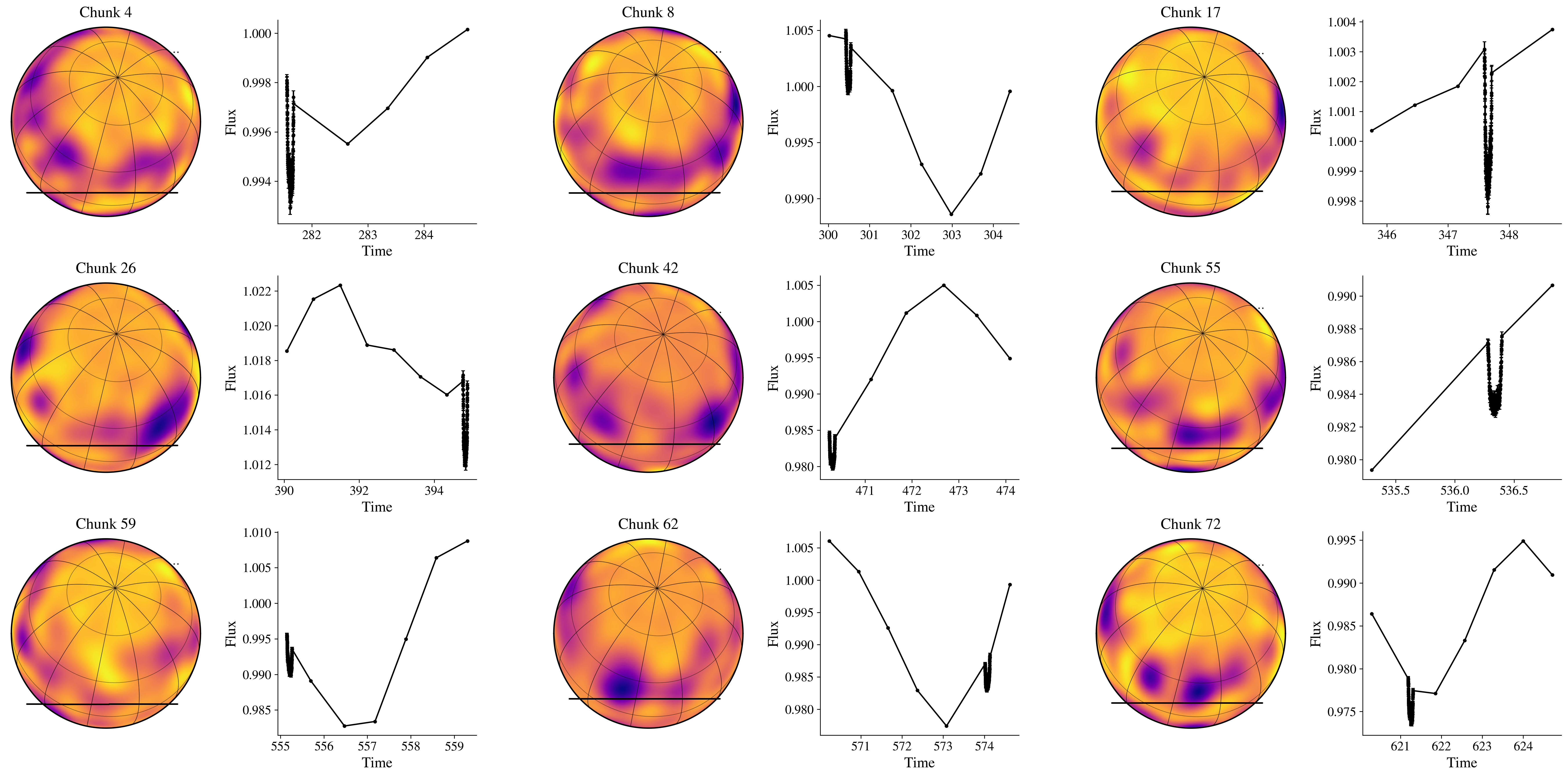}
        \caption{Temporal evolution of stellar surface maps for Kepler-63 across nine randomly-selected chunks of the 
        light curve. Each panel shows the mean stellar surface brightness distribution averaged over 
        200 MCMC samples for that specific time segment, with the black line indicating the planetary transit trajectory. 
        The color scale represents relative flux, with darker regions corresponding to cooler starspots. 
        The diversity of surface configurations demonstrates the dynamic evolution of stellar magnetic activity 
        while maintaining consistent statistical properties across all chunks.}
        \label{fig:kepler63-chunks}
    \end{centering}
\end{figure*}

To illustrate the temporal evolution of stellar surface features and their interaction with the transiting planet, 
Figure \ref{fig:kepler63-chunks} presents snapshots of the inferred stellar surface maps for nine randomly-selected chunks 
of the light curve. Each panel shows the mean stellar surface brightness distribution averaged across 200 MCMC samples for 
that specific time segment, with the black line indicating the planetary transit trajectory. The color scale represents 
the relative flux, with darker regions corresponding to cooler starspots and brighter regions representing the unspotted photosphere.
The orientation of the stellar surface maps directly reflects the high obliquity of the system, with the planetary 
trajectory cutting across the stellar disk at a steep angle relative to the stellar equator, consistent with our derived 
obliquity of $163.9^{+3.8}_{-4.2}$ degrees.
Our modeling approach treats each chunk of the light curve independently while constraining all segments to 
follow the same underlying parameter distributions. This framework allows us to capture the evolution of stellar 
surface features over time while maintaining statistical consistency across the entire dataset. The diversity of surface 
configurations shown across different chunks demonstrates the dynamic nature of the stellar magnetic field, with spots 
appearing, evolving, and disappearing on timescales comparable to the stellar rotation period.
The planetary transit chord samples different stellar latitudes depending on the specific surface configuration 
during each chunk, providing spatial information about the distribution of magnetic activity. The consistency 
of the inferred spot parameters across different chunks (as shown in the tight posterior distributions of 
Figure \ref{fig:kepler63-corener-gps}) indicates that while individual spots evolve, the overall statistical 
properties of the stellar activity remain stable over the timespan of our observations.

\section{Kepler-17}
\label{sec:kepler17}
Kepler-17 presents yet another distinct planetary system architecture that further demonstrates the versatility of 
our joint modeling approach. This system hosts a hot Jupiter, Kepler-17b, in an extremely close orbit with a period of just 
1.486 days at a distance of only 0.026 AU from its G-type main sequence host star. The planet has a mass of approximately 
2.45 $M_J$ and a radius of 1.31 $R_J$, resulting in a density of 1.35 g/cm$^3$. What makes Kepler-17 particularly 
valuable for our analysis is its intermediate activity level—more active than Kepler-45 but less extreme than Kepler-63—and 
its well-aligned orbital configuration, providing an important contrast to the highly misaligned Kepler-63 system.

\begin{figure*}[hbt!]
    \begin{centering}
        \includegraphics[width=\linewidth]{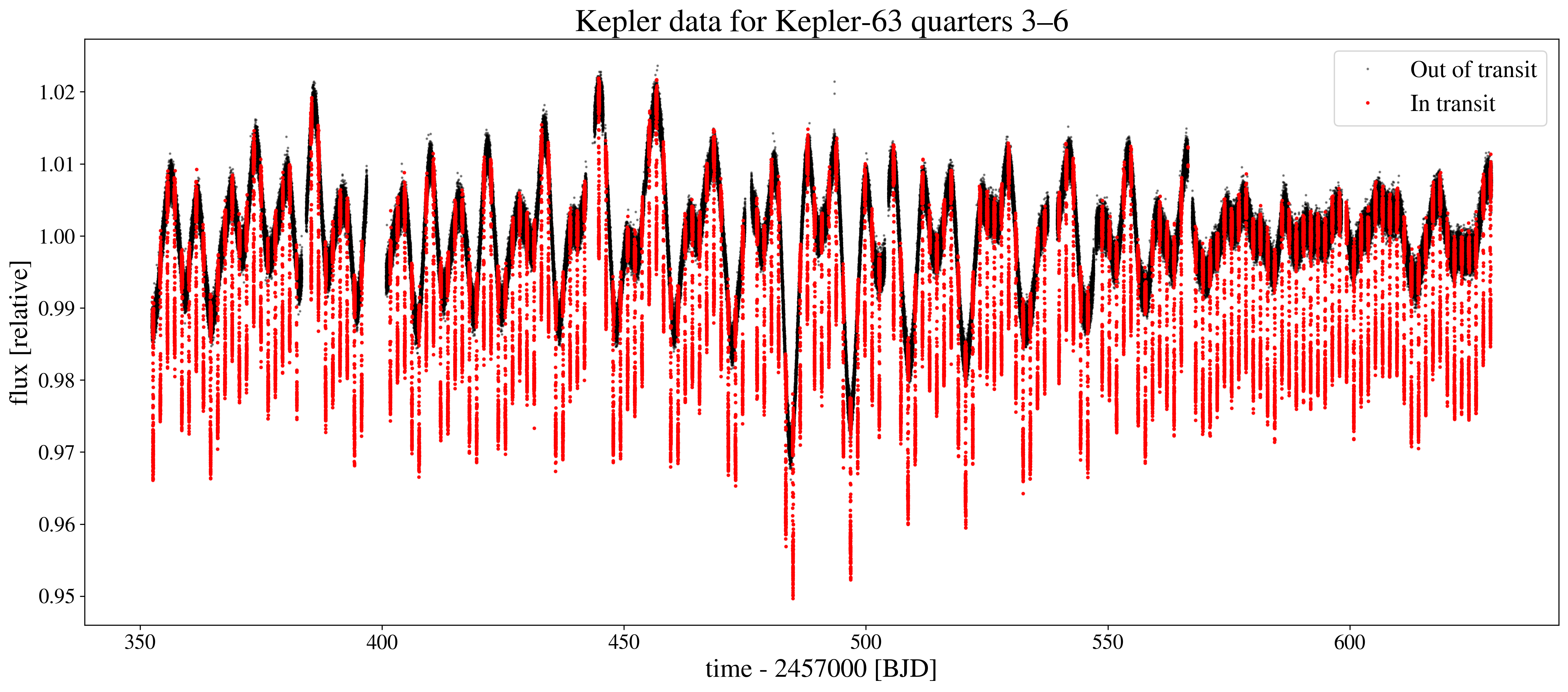}
        \caption{Complete Kepler light curve for Kepler-17 covering quarters 4, 5, 6, and 7. 
        The data spans approximately 300 days and shows frequent, deep transit events of the hot Jupiter Kepler-17b 
        against moderate stellar variability. The red points show the transits. The short 1.486-day orbital period provides excellent 
        temporal sampling of the stellar surface, while the intermediate activity level offers a valuable 
        comparison between the quiet Kepler-45 and highly active Kepler-63 systems.}
        \label{fig:kepler17-pdscap-lc}
    \end{centering}
\end{figure*}

The Kepler photometric data for Kepler-17, spanning quarters 4-7, reveals a system with substantial stellar activity 
that exhibits both the deep, frequent transits characteristic of a hot Jupiter and significant rotational modulation 
from starspots. Figure \ref{fig:kepler17-pdscap-lc} shows the complete light curve covering approximately 
300 days. The stellar variability reaches amplitudes of 2-3\%, intermediate between the moderate variations 
of Kepler-45 and the extreme activity of Kepler-63. The short orbital period results in very frequent transit events, 
providing excellent temporal sampling of the stellar surface as the planet repeatedly crosses different regions of the stellar disk.

\begin{figure*}[hbt!]
    \begin{centering}
        \includegraphics[width=\linewidth]{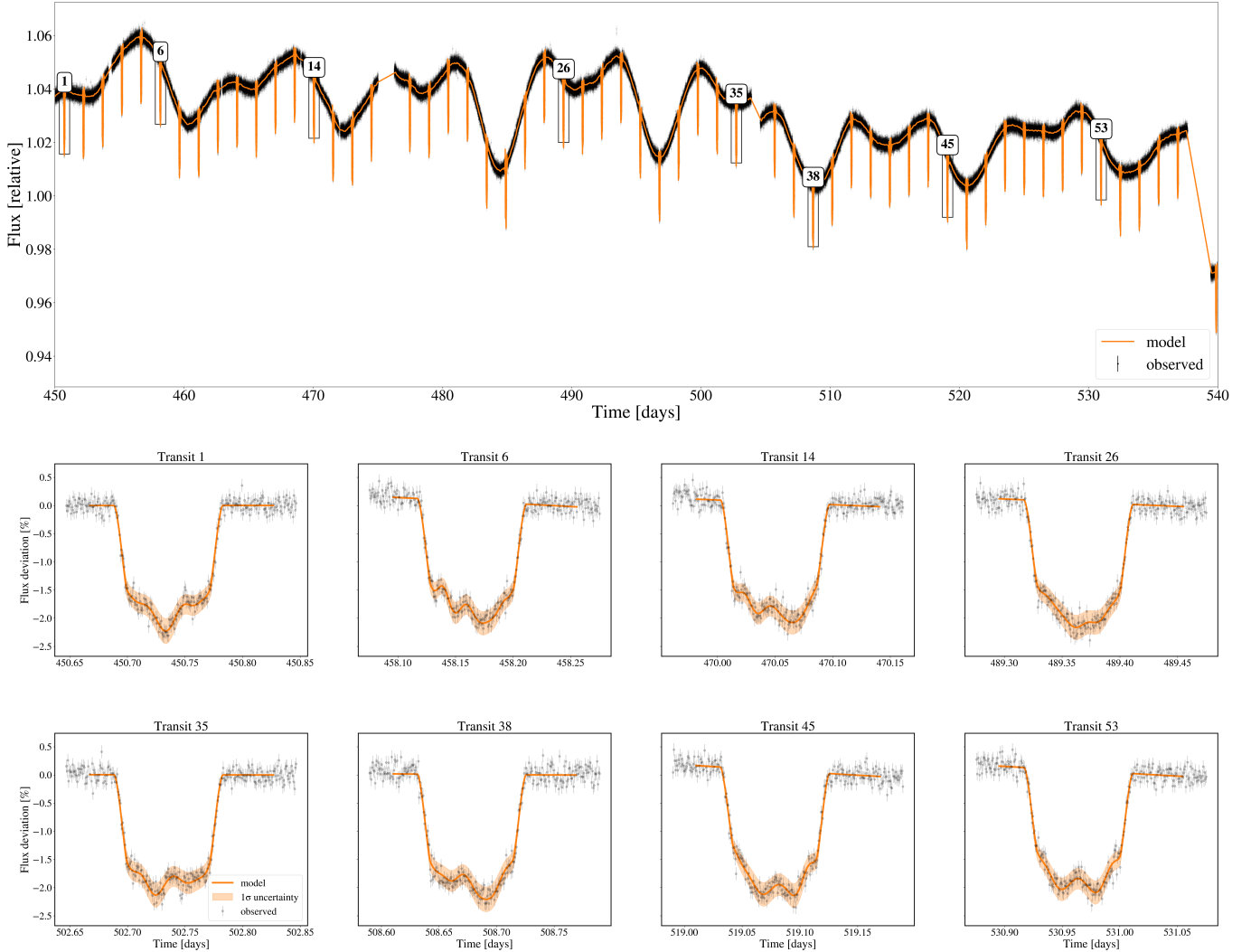}
        \caption{Transit photometry and model fits for Kepler-17. Top: Light curve segment showing frequent transit events 
        with black points representing detrended photometry and orange line showing 200 samples from the model. 
        The numbers above the boxes indicate the transit number to match the bottom panel.
        Bottom panels: Individual transit events with observed photometry (grey points with error bars) 
        and average model fits (orange lines) with $1\sigma$ uncertainty bands. The excellent fit quality despite the short 
        orbital period and moderate stellar activity demonstrates the robustness of our joint modeling approach.}
        \label{fig:kepler17-results}
    \end{centering}
\end{figure*}

Our joint modeling analysis of Kepler-17 successfully constrains both the planetary 
parameters and stellar magnetic field properties despite the challenging combination of 
short orbital period and moderate stellar activity. Figure \ref{fig:kepler17-results} demonstrates the 
quality of our fits to a representative segment of the light curve. The orange curve in the upper panel 
shows the detrended photometry, while black points represent the binned model predictions. 
The lower panels display individual transits with their model fits, showing good agreement between observations 
(grey points) and our model predictions (purple lines with uncertainty bands). The frequency of transits 
in this system provides exceptional constraints on the rotational evolution of surface features.

\begin{figure*}[hbt!]
    \begin{centering}
        \includegraphics[width=\linewidth]{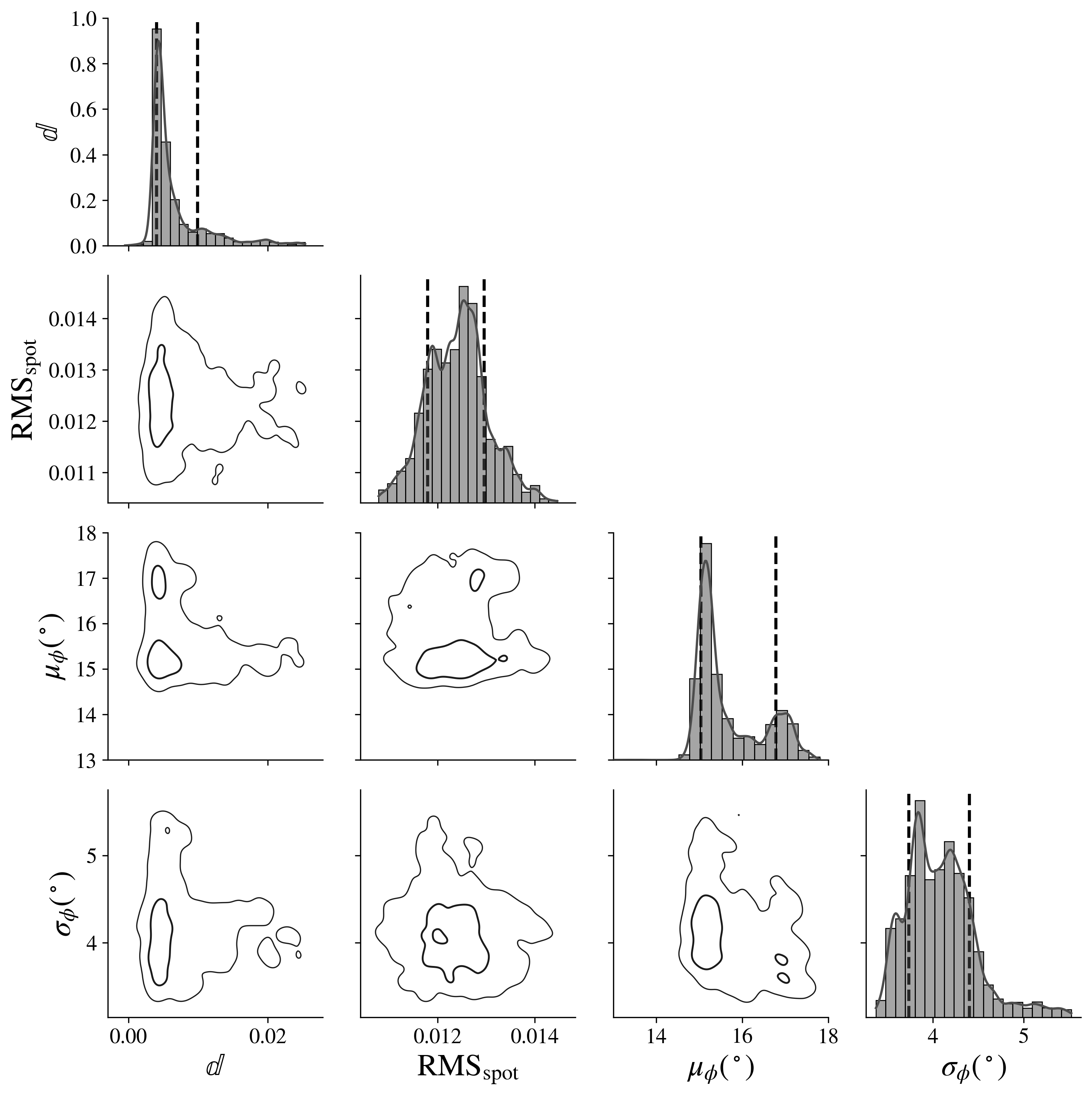}
        \caption{Posterior distributions and parameter correlations for stellar spot properties of Kepler-17. 
        Diagonal panels show marginal distributions for the flux decrement ($\mathbb{d}$), $\rm{RMS}_{\rm{spot}}$, and latitudinal distribution parameters ($\mu_\phi$ and $\sigma_\phi$). 
        Diagonal panels show marginalized posterior distributions; 
        dashed vertical lines mark the 16th and 84th percentiles ($1\sigma$ credible intervals). 
        Off-diagonal panels show two-dimensional kernel density estimates with three iso-density contours 
        illustrating correlations between parameters. The analysis reveals high-contrast spots 
        with well-constrained latitudinal preferences, reflecting the intermediate activity level and unique magnetic 
        field configuration of this system.}
        \label{fig:kepler17-corener-gps}
    \end{centering}
\end{figure*}

The stellar magnetic field characterization for Kepler-17 reveals parameter distributions that reflect 
the intermediate activity level of this system. Figure \ref{fig:kepler17-corener-gps} presents the posterior 
distributions for the starspot parameters, showing well-constrained solutions despite the complexity of the short-period 
system. Our analysis yields a spot radius of $10.09^{+0.06}_{-0.05}$ degrees (see Table \ref{tab:ResultsKepler17}), which, again,
is consistent with hitting the limits of our prior as in Section \ref{sec:kepler63} (also see Section \ref{sec:discussion} for
more discussion on this). 
Our joint modeling approach, which incorporates both rotational variability and transit constraints, favors the interpretation 
of fewer, more pronounced magnetic features. This suggests that Kepler-17 hosts concentrated, high-intensity active 
regions rather than the distributed, lower-contrast activity typical of other moderately active stars, consistent 
with the complex spot distributions identified by \cite{Namekata2020}.
The posterior distribution for $\mathbb{d}$ peaks near zero with values ranging from approximately 0.0 to 0.025. 
This indicates that starspots produce a very modest average flux decrement of approximately 0.2\%, suggesting relatively 
low levels of stellar activity. The exponential-like tail toward higher values reflects the uncertainty in constraining 
small spot coverage levels, where the posterior naturally allows for the possibility of somewhat higher activity states 
while also favoring minimal spot coverage. $\rm{RMS}_{\rm spot}$ shows a distribution centered around 0.0125. The RMS value 
of $\sim 1.25\%$ represents the typical amplitude of flux variations caused by the rotation of starspots across the 
stellar disk. 

\begin{figure*}[hbt!]
    \begin{centering}
        \includegraphics[width=\linewidth]{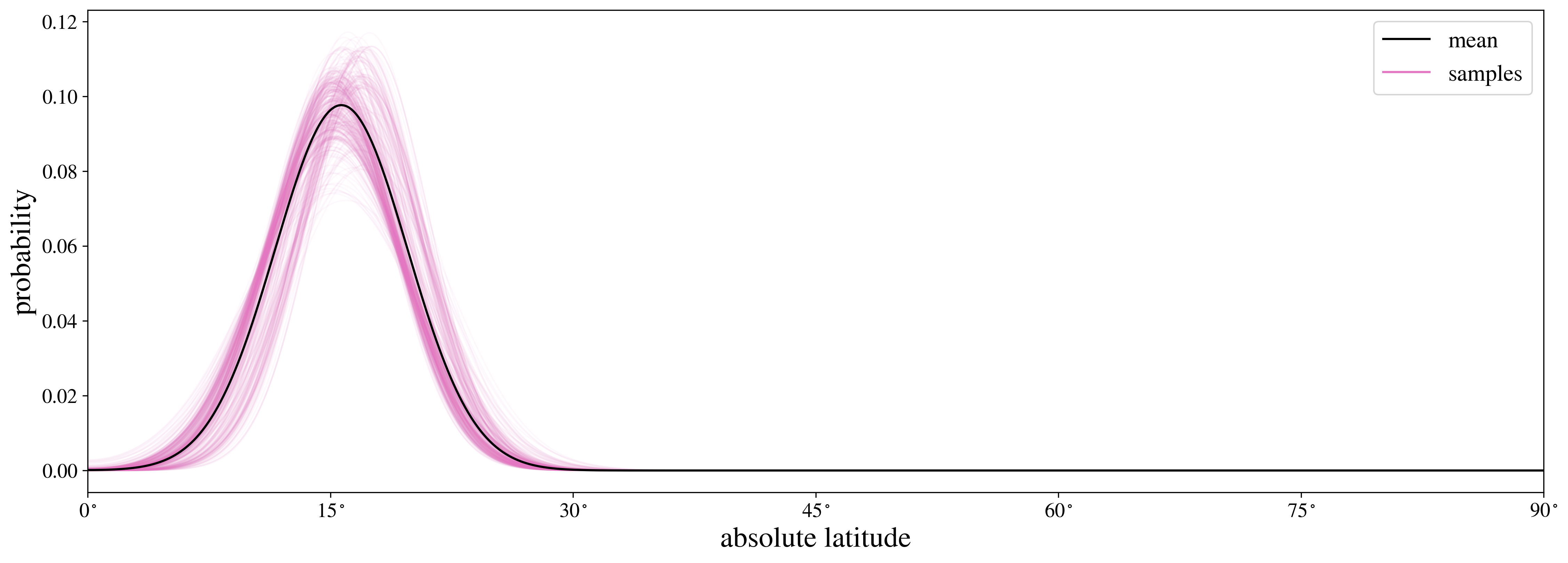}
        \caption{Latitudinal distribution of stellar activity for Kepler-17 derived from spot modeling. 
        The plot shows Posterior distributions of the absolute active latitude, shown from $0^\circ$ to $90^\circ$. 
        Thin colored curves indicate individual posterior samples, while the thick black curve shows the mean distribution.}
        \label{fig:kepler17-activelats}
    \end{centering}
\end{figure*}

The latitudinal distribution of stellar activity in Kepler-17 exhibits yet another distinct pattern, 
further illustrating the diversity of magnetic field topologies across different stellar types and activity levels. 
Figure \ref{fig:kepler17-activelats} shows a non-equatorial structure similar to Kepler-63, but with the activity bands 
located at different latitudes. The probability density reveals two symmetric peaks centered at approximately $\pm 15^\circ$, 
closer to the equator than the $\pm 30^\circ$ bands observed in Kepler-63. The black curve represents the mean posterior distribution, while the pink 
curves show individual MCMC samples, confirming the robustness of this non-equatorial pattern. The narrow latitudinal 
spread ($\sigma_\phi = 4.09^{+1.9}_{-1.7}$ degrees) indicates well-defined, concentrated activity zones, suggesting a 
more organized magnetic field structure than the broader distributions seen in the other systems.

\begin{figure*}[hbt!]
    \begin{centering}
        \includegraphics[width=\linewidth]{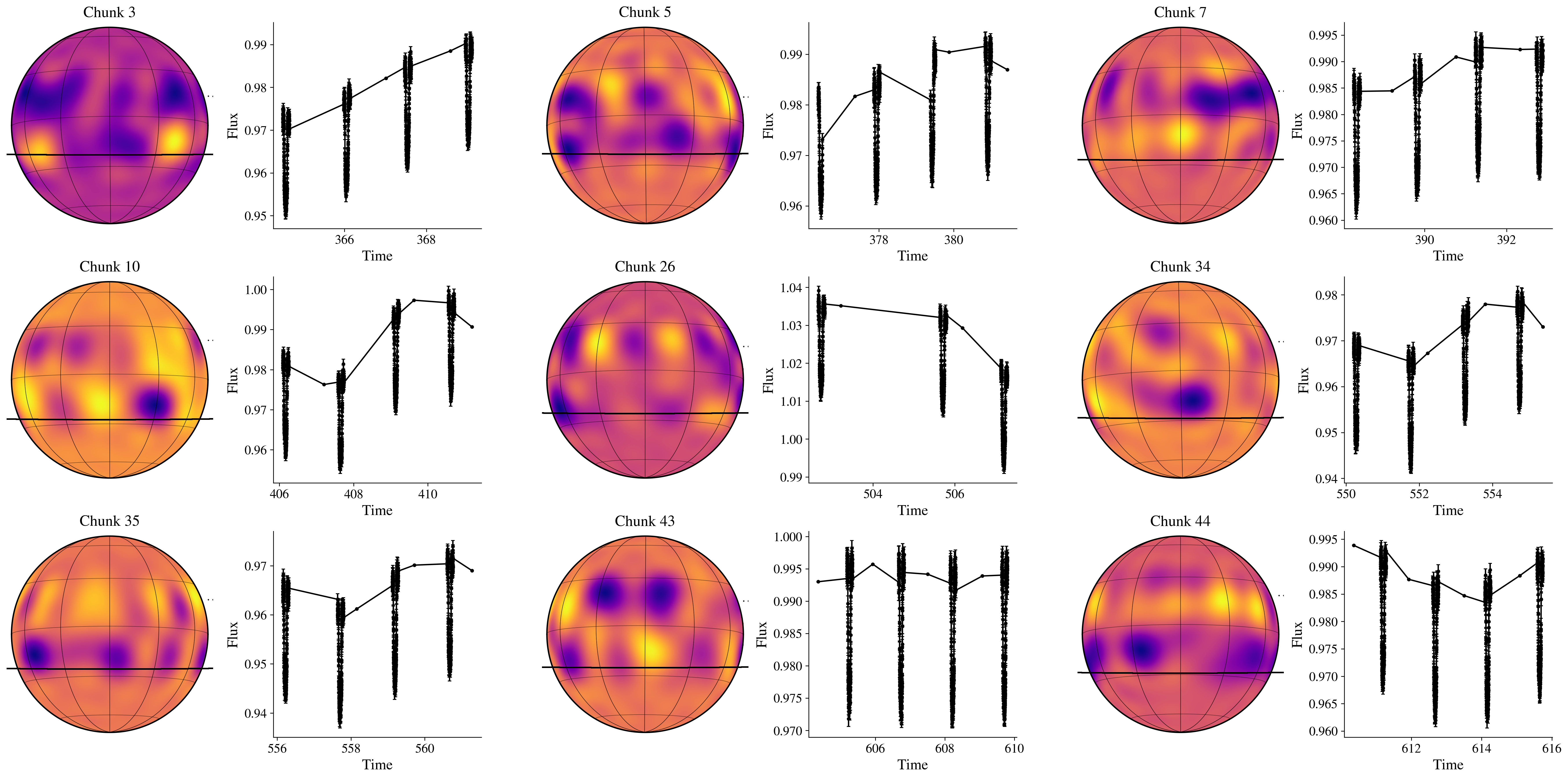}
        \caption{Temporal evolution of stellar surface maps for Kepler-17 across nine representative, randomly-selected
        light curve 
        chunks. Each panel shows the mean stellar surface brightness distribution averaged over 200 MCMC samples, 
        with the black line indicating the planetary transit trajectory. The horizontal transit path confirms the 
        well-aligned geometry of this system. The surface maps show the latitudinal activity pattern and 
        high-contrast spots characteristic of this intermediate-activity system, with clear evolution of magnetic 
        features over time.}
        \label{fig:kepler17-chunks}
    \end{centering}
\end{figure*}

The temporal evolution of the stellar surface in Kepler-17 showcases the rapid changes possible in an intermediate-activity 
system with frequent planetary transits. Figure \ref{fig:kepler17-chunks} presents snapshots of the inferred surface 
maps across nine randomly-selected chunks, with each panel showing the mean surface brightness distribution averaged over 
200 MCMC samples. The black line indicating the planetary transit trajectory appears nearly horizontal across the stellar 
disk, visually confirming the well-aligned nature of this system with our derived obliquity of $0.23^{+0.35}_{-0.43}$ degrees. 
The surface configurations show moderate variability between chunks, with clear evidence of the latitudinal activity 
pattern manifesting as enhanced spot concentrations in symmetric bands above and below the equator. The high contrast of 
individual spots is evident in several panels, where dark regions create significant brightness variations across the 
stellar surface.

Validation of our Kepler-17 analysis comes through comparison with previous detailed studies of this well-characterized system. 
Table \ref{tab:ResultsKepler17} provides a comprehensive comparison of our derived parameters with published literature 
values, demonstrating excellent agreement across all fundamental system properties. Our orbital inclination measurement 
of $86.2^{+0.93}_{-0.43}$ degrees is consistent with the literature value of $87.2\pm0.15$ degrees, while our planet-to-star 
radius ratio of $0.13 \pm 3\times10^{-4}$ agrees precisely with the published determination. Our quantitative 
obliquity measurement of $0.23^{+0.35}_{-0.43}$ degrees provides strong confirmation of the previously estimated well-aligned 
nature of this system, with our value being consistent with perfect alignment within uncertainties.

Our stellar rotation period of $11.6\pm0.1$ days matches well with the published value of $11.80\pm 1.1$ days, 
while our projected stellar rotation velocity shows excellent agreement with spectroscopic measurements. 

\begin{table}[]
    \vspace{0.5cm}
    \centering
    \caption{Comparison of inferred parameters for Kepler-17 between our analysis and literature values from \cite{Desert2011}. 
    The table shows excellent agreement for fundamental system properties while providing new constraints on stellar magnetic activity parameters.}
    \begin{tabular}{lll}
    \hline
    Parameter                                 & This Work                                          & \cite{Desert2011} \\ \hline\hline
    $i_p (^\circ)$                            & $86.2^{+0.93}_{-0.43}$                             & $87.2\pm0.15$                    \\
    $e$                                       & $0.0077^{+0.1}_{-0.006}$                           & $<0.011$                     \\
    $P (\rm days)$                            & $1.4857^{+2.3\times10^{-7}}_{-2.4\times10^{-7}}$   & $1.4857\pm 0.0000002$                   \\
    $t_0 (\rm days)$                          & $352.67\pm 2\times 10^{-5}$                        &    -                 \\
    $R_p / R_\star$                           & $0.13 \pm 3\times10^{-4}$                          &  $0.13^{+0.00022}_{-0.00018}$                   \\ 
    $u_1$                                     & $0.423^{+0.015}_{-0.011}$                          &  $0.405 \pm 0.007$                   \\ 
    $u_2$                                     & $0.156^{+0.032}_{-0.039}$                          &  $0.262^{+0.013}_{-0.015}$                   \\ 
    $v\sin i (\rm kms^{-1})$                  & $4.277\pm0.04$                                     &  $4.7\pm1.0$                   \\ \hline
    $i_\star (^\circ)$                        & $84.97^{+1.38}_{-1.71}$                            &  $\sim90$                    \\
    $\psi_\star (^\circ)$                     & $0.23^{+0.35}_{-0.43}$                             &  $0\pm15$                   \\
    $P_\star (\rm days)$                      & $11.6\pm0.1$                                       & $11.80\pm 1.1$                    \\ 
    $\rho_\star (\rm gcm^{-3})$               & $4.277\pm0.04$                                     & $4.7\pm 1$                    \\ \hline
    $\mathbb{r} (^\circ)$                     & $10.09^{+0.06}_{-0.05}$                            &  -                   \\
    $\mathbb{d}$                              & $0.0048^{+0.005}_{-0.0007}$                        &  -                   \\
    $\rm RMS_{\rm spot}$                      & $0.012^{+0.0005}_{-0.0006}$                        &  -                   \\
    $\mu_\phi (^\circ)$                       & $15.29^{+1.49}_{-0.25}$                            &  -                   \\
    $\sigma_\phi (^\circ)$                    & $4.04^{+0.36}_{-0.31}$                               &  -                   \\ \hline
    \end{tabular}
    \vspace{0.3cm}
    
    \label{tab:ResultsKepler17}
\end{table}

Our joint modeling analysis of Kepler-17 provides new insights into the stellar magnetic field topology and starspot properties 
of this moderately active solar-type star, both confirming previous measurements and revealing novel characteristics.
Our derived fundamental system parameters show excellent agreement with previous studies, providing confidence in 
our modeling approach. 

We find a clear latitudinal distribution of stellar activity, with symmetric peaks centered at approximately 
$\pm 15^\circ$ latitude. This represents the first detailed characterization of latitudinal activity structure for Kepler-17 
and reveals a magnetic field topology distinct from both the Sun and other well-studied active stars. Previous studies 
acknowledged the difficulty of constraining latitudinal information from rotational modulation alone \citep{Bonomo2012}. 
Our joint modeling approach, enabled by the precise obliquity measurement, overcomes this limitation to resolve the 
latitudinal structure. 

\section{Discussion}
\label{sec:discussion}
Our analysis of Kepler-63 and Kepler-17 reveals some evidence into the diverse
manifestations of stellar magnetic activity across stars of different spectral types, ages, and
system architectures. While the \texttt{StarryStarryProcess} framework yields robust constraints
on latitudinal distributions in each system.

\subsection{Symmetry in the Latitude Distribution}
\label{subsec:bimodality}

Our inference procedure adopts a symmetric parameterization for the spot-latitude
distribution. Specifically, we assume that spot latitudes are drawn from a
distribution $p(\phi)$ that is symmetric about the equator, which produces an
apparent bimodality in $\phi$ by construction. The symmetry in the latitude distribution follows the formulation in 
\cite{Luger2021b}, where they adopted a symmetric Beta distribution in $\cos\phi$ as a prior on spot latitudes. 
Disk-integrated photometry cannot reliably determine whether surface features lie
in the northern or southern hemisphere, so any inferred hemispheric asymmetry would
be dominated by the prior rather than by the data.

The inclination-obliquity degeneracy, though, is a \textit{physical} degeneracy in
the light-curve inversion problem arising because the stellar inclination and
hemispheric placement of features can compensate for one another. In contrast, the
bimodality in the latitude distribution arises solely from the specific form of the
symmetric prior and does not represent a constraint from the data.

Additionally, the posterior predictive latitude distributions and reconstructed surface maps
shown in this work should therefore be interpreted as \textit{statistical averages}
over the ensemble of allowed surface realizations, not as snapshots of the stellar
surface at a specific time. The maps reflect the mean properties of the spot
population inferred across each chunk; similarly, the latitude posteriors describe
the average active-latitude distribution compatible with the data under the assumed
symmetric prior.

Even with our imposed symmetry, the data provides several robust and
model-insensitive constraints. In particular:

\begin{enumerate}
    \item If spots were located at or near the equator, they would produce distinctive
    signals during transit ingress/egress or generate
    characteristic low-latitude rotational modulations. The absence of such features
    indicates a lack of equatorial activity.

    \item Rotational and transit modulation strongly constrains the longitudinal sizes
    of active regions, and these constraints remain robust even when the latitude
    distribution is symmetrized.

    \item Spots are modeled as spherical, and therefore their longitudinal and latitudinal
    extents are tied. Thus, the data-constrained longitudinal widths place indirect
    constraints on the effective latitudinal widths as well.
\end{enumerate}

\subsection{Activity Levels and Spot Properties}
The two systems span a continuum of magnetic activity. Kepler-63, the youngest and most active
target in our sample, exhibits large-amplitude rotational modulation, frequent spot evolution, and
a well-defined active-latitude spot distribution at $\pm30^\circ$ latitudes. Kepler-17 occupies an intermediate regime: 
its surface maps reveal
fewer but significantly higher-contrast spots, producing a active-latitude distribution at lower latitudes
($\pm15^\circ$). The mid-latitude belts observed in Kepler-63 and
Kepler-17 resemble solar active-latitude patterns but are shifted to higher or lower latitudes,
respectively. 

Our inferred spot properties are broadly consistent with the earlier Kepler studies, 
while also highlighting method-dependent differences that bear on interpretation.  
For Kepler-17, \cite{Lanza2019} used a maximum-entropy reconstruction of the long-cadence 
out-of-transit light curve and found persistent active longitudes.
For Kepler-63, \cite{Netto2019} used transit occultation mapping of 150 short-cadence transits 
to measure the spot latitudes directly, and reported a clear bimodal latitude distribution 
with a gap near $\sim$34$^\circ$.
Our joint \texttt{StarryStarryProcess} inferences recover the same broad phenomenology 
but with some quantitative differences that reflect modelling choices and effective resolution. 
We infer characteristic spot angular radii near $\sim$10$^\circ$ for both stars 
(Kepler-63: $10.02^{+0.03}_{-0.02}{}^\circ$; Kepler-17: $10.09^{+0.06}_{-0.05}{}^\circ$), consistent with 
the spot sizes reported in the transit-mapping literature. For Kepler-17 we recover active latitudes concentrated closer to the equator (peaks at $\sim\pm15^\circ$).
We need to keep in mind that map geometry and observables differ while comparing different results with
different techniques. \cite{Lanza2019} reconstruct longitudinal spot filling factors from 
long-cadence out-of-transit photometry and validate against occulted spots, while \cite{Netto2019} 
derive direct spot latitudes from individual occultations; our method jointly models rotational modulation and transits, 
which helps break some degeneracies but also mixes the information that each single technique isolates.

We also show two different spin-orbit orientations. The high-obliquity system, Kepler-63, reveals more 
complex or structured spot distributions, with the
planetary transit chord sampling a broad swath of stellar latitudes. By contrast, the nearly aligned system Kepler-17 
primarily reveals equatorial activity, both because of genuine magnetic preferences and
because the transit geometry restricts the probed latitudes. 

\subsection{Resolution Limitations}

A critical limitation of our analysis stems from the spherical harmonic resolution employed in the \texttt{StarryStarryProcess} 
model. With $\ell_{\rm max} = 15$, the model cannot adequately represent starspots with angular radii smaller than approximately 
$10^\circ$, a constraint that becomes particularly problematic for Sun-like stars where observed spot sizes are 
typically much smaller \citep{Baumann2005}. For example, the largest recorded sunspot reached only $6^\circ$ in angular radius, 
while typical active region complexes are considerably smaller. This resolution limit manifests as an 
effective lower bound in our posterior distributions for spot parameters, artificially constraining the model to 
favor large starspot features. The model's inability to reproduce the sharp, 
localized flux variations characteristic of small starspot occultations leads to compensatory adjustments in other parameters, 
potentially biasing our estimates of spot coverage, contrast, and spatial distribution. The posterior samples consistently 
push against the resolution limit in both Kapler-63 and Kepler-17, suggesting that the true underlying spot populations contain 
significant populations of features below our detection threshold.

The model attempts to reconcile observed flux variations with the active latitude framework by 
adjusting the spot latitude distribution parameters $\{\mu_\psi, \sigma_\psi\}$ to place the observed variability 
within the geometrically accessible region. However, when the true spot population consists of smaller 
features distributed across a broader latitude range than our model can represent, this optimization 
process may converge on biased solutions. The active latitude inference becomes dominated by the largest, 
most readily detectable features, while the contribution from numerous smaller spots may be misattributed to 
changes in the overall activity level or latitude distribution width. Potentially, these systematic effects can impact 
our obliquity measurements, as the stellar inclination-obliquity degeneracy breaking relies critically on 
accurate modeling of spot latitude distributions and their intersection with the transit geometry. If the model 
systematically misrepresents the true spot population due to resolution limitations, the resulting obliquity constraints 
may be biased toward configurations that compensate for the missing small-scale structure. Despite these resolution 
limitations, our methodology successfully recovers meaningful obliquity measurements for the systems analyzed. 
The obliquity constraints remain robust because they primarily depend on the large-scale pattern of 
spot occultations during transits and the overall rotational modulation signature, 
rather than the detailed representation of individual small spots. The geometric relationship between the 
transit chord, stellar rotation axis, and dominant spot-bearing latitudes can be 
inferred even when individual spots are not perfectly resolved. The active latitude framework captures the mean and 
dispersion of the spot distribution, parameters that remain well-constrained even when the underlying population 
contains unresolved small-scale structure. The aggregate photometric signature from many small spots can still inform the 
overall latitude distribution parameters, allowing the model to recover meaningful constraints on the bulk 
properties of stellar magnetic activity.



\section{Conclusions}

We have applied the \texttt{StarryStarryProcess} framework to archival \textit{Kepler} light curves
of two transiting exoplanet host stars: Kepler-63 and Kepler-17. This
analysis expands upon the validation of the method on TOI-3884 presented in \paperone, and
demonstrates its ability to jointly constrain planetary parameters and stellar surface magnetic
properties across diverse stellar environments.

Our main conclusions are as follows:

\begin{enumerate}
    \item Both stars show a characteristic spot scale of $\sim$10$^\circ$, which is the 
    natural resolution limit of our spherical-harmonic representation but also agrees with spot 
    sizes inferred from transit analyses.  

    \item Both stars
    exhibit clear active-latitude spot distributions, with active latitudes at $\pm 30^\circ$ and $\pm 15^\circ$,
    respectively. 

    \item The misaligned system Kepler-63 reveals more complex or structured spot distributions, while the aligned system
    Kepler-17 shows primarily close to equatorial activity. This may partly reflect
    intrinsic stellar differences, but also illustrates the observational bias introduced by
    transit geometry.

    \item We were able to constrain stellar obliquity and inclinations using one-band photometry and compare them
    with the literature spot-induced or Rossiter-Maclaughlin measurements, which validates our approach.
\end{enumerate}

This comparative framework demonstrates the power of transit photometry, when combined with
Bayesian surface-mapping tools, to probe stellar magnetism beyond the Sun. With larger samples
from \textit{TESS} and future missions such as \textit{PLATO}, population-level studies will enable
direct tests of dynamo theory across stellar types and evolutionary stages, and clarify the role of
stellar obliquity in shaping both magnetic activity and planetary system architectures.

\section*{Acknowledgments}
We thank Brett Morris for extensive comments that significantly improved the paper and our understanding of the 
influence of and interplay between the $\mathbb{n}$ , $\mathbb{c}$, and $\mathbb{r}$ variables. We also thank the anonymous referee
for the useful comments.
S.S. thanks the LSST-DA Data Science Fellowship Program, which is funded by LSST-DA, the Brinson Foundation, 
the WoodNext Foundation, and the Research Corporation for Science Advancement Foundation; 
her participation in the program has benefited this work. S.S. acknowledges support from award 644616 from the Simons Foundation.

\bibliography{bib}

\begin{thebibliography}{}
\expandafter\ifx\csname natexlab\endcsname\relax\def\natexlab#1{#1}\fi
\providecommand{\url}[1]{\href{#1}{#1}}
\providecommand{\dodoi}[1]{doi:~\href{http://doi.org/#1}{\nolinkurl{#1}}}
\providecommand{\doeprint}[1]{\href{http://ascl.net/#1}{\nolinkurl{http://ascl.net/#1}}}
\providecommand{\doarXiv}[1]{\href{https://arxiv.org/abs/#1}{\nolinkurl{https://arxiv.org/abs/#1}}}

\bibitem[{{Baumann} \& {Solanki}(2005)}]{Baumann2005}
{Baumann}, I., \& {Solanki}, S.~K. 2005, \aap, 443, 1061,
  \dodoi{10.1051/0004-6361:20053415}

\bibitem[{{B{\'e}ky} {et~al.}(2014){B{\'e}ky}, {Kipping}, \&
  {Holman}}]{Beky2014}
{B{\'e}ky}, B., {Kipping}, D.~M., \& {Holman}, M.~J. 2014, \mnras, 442, 3686,
  \dodoi{10.1093/mnras/stu1061}

\bibitem[{{Bonomo} \& {Lanza}(2012)}]{Bonomo2012}
{Bonomo}, A.~S., \& {Lanza}, A.~F. 2012, \aap, 547, A37,
  \dodoi{10.1051/0004-6361/201219999}

\bibitem[{{Borucki} {et~al.}(2010){Borucki}, {Koch}, {Basri}, {Batalha},
  {Brown}, {Caldwell}, {Caldwell}, {Christensen-Dalsgaard}, {Cochran},
  {DeVore}, {Dunham}, {Dupree}, {Gautier}, {Geary}, {Gilliland}, {Gould},
  {Howell}, {Jenkins}, {Kondo}, {Latham}, {Marcy}, {Meibom}, {Kjeldsen},
  {Lissauer}, {Monet}, {Morrison}, {Sasselov}, {Tarter}, {Boss}, {Brownlee},
  {Owen}, {Buzasi}, {Charbonneau}, {Doyle}, {Fortney}, {Ford}, {Holman},
  {Seager}, {Steffen}, {Welsh}, {Rowe}, {Anderson}, {Buchhave}, {Ciardi},
  {Walkowicz}, {Sherry}, {Horch}, {Isaacson}, {Everett}, {Fischer}, {Torres},
  {Johnson}, {Endl}, {MacQueen}, {Bryson}, {Dotson}, {Haas}, {Kolodziejczak},
  {Van Cleve}, {Chandrasekaran}, {Twicken}, {Quintana}, {Clarke}, {Allen},
  {Li}, {Wu}, {Tenenbaum}, {Verner}, {Bruhweiler}, {Barnes}, \&
  {Prsa}}]{Borucki2010}
{Borucki}, W.~J., {Koch}, D., {Basri}, G., {et~al.} 2010, Science, 327, 977,
  \dodoi{10.1126/science.1185402}

\bibitem[{{D{\'e}sert} {et~al.}(2011){D{\'e}sert}, {Charbonneau}, {Demory},
  {Ballard}, {Carter}, {Fortney}, {Cochran}, {Endl}, {Quinn}, {Isaacson},
  {Fressin}, {Buchhave}, {Latham}, {Knutson}, {Bryson}, {Torres}, {Rowe},
  {Batalha}, {Borucki}, {Brown}, {Caldwell}, {Christiansen}, {Deming},
  {Fabrycky}, {Ford}, {Gilliland}, {Gillon}, {Haas}, {Jenkins}, {Kinemuchi},
  {Koch}, {Lissauer}, {Lucas}, {Mullally}, {MacQueen}, {Marcy}, {Sasselov},
  {Seager}, {Still}, {Tenenbaum}, {Uddin}, \& {Winn}}]{Desert2011}
{D{\'e}sert}, J.-M., {Charbonneau}, D., {Demory}, B.-O., {et~al.} 2011, \apjs,
  197, 14, \dodoi{10.1088/0067-0049/197/1/14}

\bibitem[{{Ducrot} {et~al.}(2018){Ducrot}, {Sestovic}, {Morris}, {Gillon},
  {Triaud}, {De Wit}, {Thimmarayappa}, {Agol}, {Almleaky}, {Burdanov},
  {Burgasser}, {Delrez}, {Demory}, {Jehin}, {Leconte}, {McCormac}, {Murray},
  {Queloz}, {Selsis}, {Thompson}, \& {Van Grootel}}]{Ducrot2018}
{Ducrot}, E., {Sestovic}, M., {Morris}, B.~M., {et~al.} 2018, \aj, 156, 218,
  \dodoi{10.3847/1538-3881/aade94}

\bibitem[{{Herrero} {et~al.}(2016){Herrero}, {Ribas}, {Jordi}, {Morales},
  {Perger}, \& {Rosich}}]{Herrero2016}
{Herrero}, E., {Ribas}, I., {Jordi}, C., {et~al.} 2016, \aap, 586, A131,
  \dodoi{10.1051/0004-6361/201425369}

\bibitem[{{Huber} {et~al.}(2010){Huber}, {Czesla}, {Wolter}, \&
  {Schmitt}}]{Huber2010}
{Huber}, K.~F., {Czesla}, S., {Wolter}, U., \& {Schmitt}, J.~H.~M.~M. 2010,
  \aap, 514, A39, \dodoi{10.1051/0004-6361/200913914}

\bibitem[{{Juvan} {et~al.}(2018){Juvan}, {Lendl}, {Cubillos}, {Fossati},
  {Tregloan-Reed}, {Lammer}, {Guenther}, \& {Hanslmeier}}]{Juvan2018}
{Juvan}, I.~G., {Lendl}, M., {Cubillos}, P.~E., {et~al.} 2018, \aap, 610, A15,
  \dodoi{10.1051/0004-6361/201731345}

\bibitem[{{Lanza} {et~al.}(2019){Lanza}, {Netto}, {Bonomo}, {Parviainen},
  {Valio}, \& {Aigrain}}]{Lanza2019}
{Lanza}, A.~F., {Netto}, Y., {Bonomo}, A.~S., {et~al.} 2019, \aap, 626, A38,
  \dodoi{10.1051/0004-6361/201833894}

\bibitem[{{Lim} {et~al.}(2023){Lim}, {Benneke}, {Doyon}, {MacDonald},
  {Piaulet}, {Artigau}, {Coulombe}, {Radica}, {L'Heureux}, {Albert}, {Rackham},
  {de Wit}, {Salhi}, {Roy}, {Flagg}, {Fournier-Tondreau}, {Taylor}, {Cook},
  {Lafreni{\`e}re}, {Cowan}, {Kaltenegger}, {Rowe}, {Espinoza}, {Dang}, \&
  {Darveau-Bernier}}]{Lim2023}
{Lim}, O., {Benneke}, B., {Doyon}, R., {et~al.} 2023, \apjl, 955, L22,
  \dodoi{10.3847/2041-8213/acf7c4}

\bibitem[{{Luger} {et~al.}(2019){Luger}, {Agol}, {Foreman-Mackey}, {Fleming},
  {Lustig-Yaeger}, \& {Deitrick}}]{Luger2019}
{Luger}, R., {Agol}, E., {Foreman-Mackey}, D., {et~al.} 2019, \aj, 157, 64,
  \dodoi{10.3847/1538-3881/aae8e5}

\bibitem[{{Luger} {et~al.}(2021{\natexlab{a}}){Luger}, {Foreman-Mackey}, \&
  {Hedges}}]{Luger2021b}
{Luger}, R., {Foreman-Mackey}, D., \& {Hedges}, C. 2021{\natexlab{a}}, \aj,
  162, 124, \dodoi{10.3847/1538-3881/abfdb9}

\bibitem[{{Luger} {et~al.}(2021{\natexlab{b}}){Luger}, {Foreman-Mackey},
  {Hedges}, \& {Hogg}}]{Luger2021a}
{Luger}, R., {Foreman-Mackey}, D., {Hedges}, C., \& {Hogg}, D.~W.
  2021{\natexlab{b}}, \aj, 162, 123, \dodoi{10.3847/1538-3881/abfdb8}

\bibitem[{{Maxted}(2016)}]{Maxted2016}
{Maxted}, P.~F.~L. 2016, \aap, 591, A111, \dodoi{10.1051/0004-6361/201628579}

\bibitem[{{Montalto} {et~al.}(2014){Montalto}, {Bou{\'e}}, {Oshagh}, {Boisse},
  {Bruno}, \& {Santos}}]{Montalto2014}
{Montalto}, M., {Bou{\'e}}, G., {Oshagh}, M., {et~al.} 2014, \mnras, 444, 1721,
  \dodoi{10.1093/mnras/stu1530}

\bibitem[{{Morris} {et~al.}(2018){Morris}, {Agol}, {Hebb}, \&
  {Hawley}}]{Morris2018}
{Morris}, B.~M., {Agol}, E., {Hebb}, L., \& {Hawley}, S.~L. 2018, \aj, 156, 91,
  \dodoi{10.3847/1538-3881/aad3b7}

\bibitem[{{Namekata} {et~al.}(2020){Namekata}, {Davenport}, {Morris}, {Hawley},
  {Maehara}, {Notsu}, {Toriumi}, {Ikuta}, {Notsu}, {Honda}, {Nogami}, \&
  {Shibata}}]{Namekata2020}
{Namekata}, K., {Davenport}, J. R.~A., {Morris}, B.~M., {et~al.} 2020, \apj,
  891, 103, \dodoi{10.3847/1538-4357/ab7384}

\bibitem[{{Netto, Y.} \& {Valio, A.}(2020)}]{Netto2019}
{Netto, Y.}, \& {Valio, A.} 2020, A\&A, 635, A78,
  \dodoi{10.1051/0004-6361/201936219}

\bibitem[{{Oshagh} {et~al.}(2013){Oshagh}, {Boisse}, {Bou{\'e}}, {Montalto},
  {Santos}, {Bonfils}, \& {Haghighipour}}]{Oshagh2013}
{Oshagh}, M., {Boisse}, I., {Bou{\'e}}, G., {et~al.} 2013, \aap, 549, A35,
  \dodoi{10.1051/0004-6361/201220173}

\bibitem[{{Sagynbayeva} {et~al.}(2025){Sagynbayeva}, {Farr}, {Morris}, \&
  {Luger}}]{paper1}
{Sagynbayeva}, S., {Farr}, W.~M., {Morris}, B.~M., \& {Luger}, R. 2025, \apj,
  990, 32, \dodoi{10.3847/1538-4357/adf6be}

\bibitem[{{Sanchis-Ojeda} \& {Winn}(2011)}]{Sanchis2011}
{Sanchis-Ojeda}, R., \& {Winn}, J.~N. 2011, \apj, 743, 61,
  \dodoi{10.1088/0004-637X/743/1/61}

\bibitem[{{Sanchis-Ojeda} {et~al.}(2013){Sanchis-Ojeda}, {Winn}, {Marcy},
  {Howard}, {Isaacson}, {Johnson}, {Torres}, {Albrecht}, {Campante}, {Chaplin},
  {Davies}, {Lund}, {Carter}, {Dawson}, {Buchhave}, {Everett}, {Fischer},
  {Geary}, {Gilliland}, {Horch}, {Howell}, \& {Latham}}]{Sanchis2013}
{Sanchis-Ojeda}, R., {Winn}, J.~N., {Marcy}, G.~W., {et~al.} 2013, \apj, 775,
  54, \dodoi{10.1088/0004-637X/775/1/54}

\bibitem[{{Scandariato} {et~al.}(2017){Scandariato}, {Nascimbeni}, {Lanza},
  {Pagano}, {Zanmar Sanchez}, \& {Leto}}]{Scandariato2017}
{Scandariato}, G., {Nascimbeni}, V., {Lanza}, A.~F., {et~al.} 2017, \aap, 606,
  A134, \dodoi{10.1051/0004-6361/201730966}

\bibitem[{{Silva}(2003)}]{Silva2003}
{Silva}, A. V.~R. 2003, \apjl, 585, L147, \dodoi{10.1086/374324}

\bibitem[{{Tregloan-Reed} {et~al.}(2013){Tregloan-Reed}, {Southworth}, \&
  {Tappert}}]{Tregloan-Reed2013}
{Tregloan-Reed}, J., {Southworth}, J., \& {Tappert}, C. 2013, \mnras, 428,
  3671, \dodoi{10.1093/mnras/sts306}

\bibitem[{{Valio} {et~al.}(2017){Valio}, {Estrela}, {Netto}, {Bravo}, \& {de
  Medeiros}}]{Valio2017}
{Valio}, A., {Estrela}, R., {Netto}, Y., {Bravo}, J.~P., \& {de Medeiros},
  J.~R. 2017, \apj, 835, 294, \dodoi{10.3847/1538-4357/835/2/294}

\end{thebibliography}
\end{document}